\documentstyle[emulateapj,apjfonts,psfig,natbib,amsmath]{article}

\begin{document}

\def\cit{1}
\def\tapir{2}
\def\ociw{3}
\def\pu{4}
\def\hf{5}

\title{Late-time Radio Observations of 68 Type Ibc Supernovae: \\
  Strong Constraints on Off-Axis Gamma-ray Bursts}

\author{A.~M. Soderberg \altaffilmark{\cit}, E. Nakar \altaffilmark{\tapir},
E. Berger \altaffilmark{\ociw,\pu,\hf}, S.~R. Kulkarni \altaffilmark{\cit}}

\altaffiltext{\cit}{Division of Physics, Mathematics and Astronomy,
        105-24, California Institute of Technology, Pasadena, CA
        91125} 
\altaffiltext{\tapir}{Theoretical Astrophysics, 130-33
        California Institute of Technology, Pasadena, CA 91125}
\altaffiltext{\ociw}{Observatories of the Carnegie Institution of Washington,
        813 Santa Barbara St., Pasadena, CA 91101}
\altaffiltext{\pu}{Department of Astrophysical Sciences, Princeton University, 
        Princeton, NJ 08544}
\altaffiltext{\hf}{Hubble Fellow}

\begin{abstract}
We present late-time radio observations of 68 local Type Ibc
supernovae, including six events with broad optical absorption lines
(``hypernovae'').  None of these objects exhibit radio
emission attributable to off-axis gamma-ray burst jets spreading into
our line-of-sight.  Comparison with our afterglow models reveals the
following conclusions: (1) Less than $\sim 10\%$ of Type Ibc
supernovae are associated with typical gamma-ray bursts initially
directed away from our line-of-sight; this places an empirical
constraint on the GRB beaming factor of $\langle f_b^{-1}\rangle
\lesssim 10^4$ corresponding to an average jet opening angle,
$\theta_j \gtrsim 0.8$ degrees. (2) This holds in particular for the
broad-lined supernovae (SNe 1997dq, 1997ef, 1998ey, 2002ap,
2002bl and 2003jd) which have been argued to host GRB jets.  Our
observations reveal no evidence for typical (or even sub-energetic)
GRBs and rule out the scenario in which every broad-lined SN harbors
a GRB at the $84\%$ confidence level.  Their large photospheric
velocities and asymmetric ejecta (inferred from
spectropolarimetry and nebular spectroscopy) appear to be
characteristic of the non-relativistic supernova explosion and do not
necessarily imply the existence of associated GRB jets.
\end{abstract}


\section{Introduction}
It is now generally accepted that long duration gamma-ray bursts
(GRBs) give rise to engine-driven relativistic jets as well as
non-relativistic spherical supernova (SN) explosions.  The first
example of this GRB-SN connection came with the discovery of the Type
Ic supernova, SN\,1998bw, associated with GRB\,980425 ($d\sim 36$ Mpc;
\citealt{gvv+98,paa+00}).  The unusually fast photospheric velocities
and exceptionally bright radio emission of SN\,1998bw indicated $\sim
10^{52}$ erg of kinetic energy and mildly relativistic ejecta (bulk
Lorentz factor, $\Gamma\sim 3$; \citealt{kfw+98,imn+98,lc99,wes99}).
In comparison with other core-collapse events ($E_{KE}\sim 10^{51}$
erg and ejecta speeds, $v\lesssim 0.1c$), SN\,1998bw was considered a
hyper-energetic supernova (``hypernova''; \citealt{imn+98}).  Broad
optical absorption lines were also observed in the Type Ic SNe 2003dh
and 2003lw, associated with the cosmological GRBs 030329 and 031203,
indicative of comparably large photospheric velocities
\citep{mgs+03,mtc+04}.  Together, these observations appear to suggest
that broad spectral features are characteristic of GRB-associated SNe.

In addition to events with prompt gamma-ray emission, the GRB-SN
connection also implies the existence of ``orphan'' supernovae whose
relativistic jets are initially beamed away from our line of sight
\citep{rho99,pac01}.  Since the discovery of SN\,1998bw, several
broad-lined SNe have been identified locally ($d\lesssim 100$ Mpc) and
are currently estimated to represent $\sim 5\%$ of the Type Ibc
supernova (SNe Ibc) population \citep{pmn+04}.  Given their spectral
similarity to the GRB-associated SNe, it has been argued that local
broad-lined supernovae can be used as signposts for GRBs.

Thus, associations with poorly-localized BATSE bursts have been
invoked for the broad-lined SNe 1997cy, 1997ef and
1999E\footnotemark\footnotetext{We note that SNe 1997cy and 1999E were
  initially classified as Type IIn supernovae while \citet{hps+03}
  later showed convincing evidence that they are hydrogen-rich Type Ia
  events similar to SN\,2002ic.}
\citep{grs+00,tsm+00,ww98,min00,rtb+03}.  In addition, association
with off-axis GRBs have also been claimed.  In the case of SN\,2002ap,
broad optical absorption lines and evidence for mildly asymmetric
ejecta (based on spectropolarimetry measurements) were interpreted to
support an off-axis GRB jet (\citealt{kji+02,tot03}, but see
\citealt{lfc+02}).

More recently, an off-axis GRB model has been proposed for SN\,2003jd,
for which photospheric velocities upward of 40,000~$\rm km~s^{-1}$
were measured at early time \citep{ffs03,mck+03}.  More intriguingly,
late-time ($t\sim 400$ days) spectra showed double-peaked emission
lines of light-elements, attributed to an asymmetric explosion
\citep{kmd+04}.  \citet{mkm+05} argue that these observations can be
understood if SN\,2003jd was accompanied by a highly collimated GRB
jet initially directed $\sim 70$ degrees away from our line-of-sight.

Regardless of viewing angle, however, strong afterglow emission
eventually becomes visible as the decelerating GRB jets spread
laterally and the emission becomes effectively isotropic.  As the jets
spread into our line-of-sight, a rapid increase of broadband
synchrotron emission is observed on a timescale of a few weeks to
several years.  This late-time emission is most easily detected at
long wavelengths \citep{pl98,low+02,wax04}.  Targeting local Type Ibc
supernovae with late-time radio observations has thus become the
preferred method to search for evidence of off-axis GRBs
\citep{svw+03,sfw04}.

Using early radio observations ($t\lesssim 100$ days) we have already
limited the fraction of SNe Ibc harboring on-axis (or mildly off-axis)
GRBs to be $\lesssim 3\%$ \citep{bkf+03}.  In this paper, we present
late-time ($t\sim 0.5$ to 20 yr) radio observations for 68 local Type
Ibc supernovae, including SN\,2003jd and five additional broad-lined
events, making this the most comprehensive study of late-time radio
emission from SNe Ibc.  We use these data to constrain the SN fraction
associated with GRB jets regardless of viewing angle assumptions,
constraining even those initially beamed perpendicular to our
line-of-sight.

\section{Radio Observations}
\label{sec:obs}

\subsection{Type Ic SN\,2003jd}

SN\,2003jd was discovered on 2003 October 25.2 UT within host galaxy
MCG -01-59-021 ($d_L\sim 81$ Mpc; \citealt{bsl+03}).  In
Table~\ref{tab:SN2003jd} we summarize our radio observations for
SN\,2003jd, spanning $8-569$ days after the
explosion\footnotemark\footnotetext{Here we assume an approximate
  explosion date of 2003 October 21 UT, based on pre-explosion images
  \citep{bsl+03}.}.  All observations were conducted with the Very
Large Array\footnotemark\footnotetext{The Very Large Array is a
  facility of the National Science Foundation operated under
  cooperative agreement by Associated Universities, Inc.} (VLA) in the
standard continuum mode with a bandwidth of $2\times 50$ MHz centered
at 4.86, 8.46 or 22.5 GHz.  We used 3C48 and 3C147 (J0137+331 and
J0542+498) for flux calibration, while J2323-032 was used to monitor
the phase.  Data were reduced using standard packages within the
Astronomical Image Processing System (AIPS).

No radio emission was detected at the optical SN position during our
early observations.  Our radio limits imply that SN\,2003jd was a
factor of $\gtrsim 100$ less luminous than SN\,1998bw on a comparable
timescale.  We conclude that SN\,2003jd, like the majority of SNe Ibc,
did not produce relativistic ejecta along our line-of-sight.

We re-observed SN\,2003jd at $t\sim 1.6$ yrs to search for radio
emission from an off-axis GRB jet.  No emission was detected, implying
a limit of $F_{\nu} < 45~\mu$Jy ($3\sigma$) at 8.46 GHz.

\subsection{Late-time data on Local Type Ibc Supernovae}

We supplement these data with late-time ($t\sim 0.5-20$ year) radio
observations for 67 local ($d_L\lesssim 200$ Mpc) SNe Ibc, summarized
in Table~\ref{tab:vla}.  Eleven objects were observed at moderately
late-time as part of our on-going VLA program to characterize the
early ($t\lesssim 100$ days) radio emission from SNe Ibc (Soderberg
{\it et al.}, in prep).  The remaining 54 objects were observed on a
later timescale ($t\gtrsim 1$ year) and were taken from the VLA
archive\footnotemark\footnotetext{http://e2e.nrao.edu/archive/}.  We
note that five of these supernovae (SNe 1997dq, 1997ef, 1998ey,
2002ap, 2002bl) were spectroscopically observed to have broad optical
absorption lines, similar to SN\,1998bw.

All VLA observations were conducted at 8.46 GHz (except for SN\,1991D
at 4.86 GHz) in the standard continuum mode with a bandwidth of
$2\times 50$ MHz.  Data were reduced using AIPS, and the resulting
flux density measurements for this sample of SNe Ibc is given in
Table~\ref{tab:vla}.  With the exception of SN\,2001em, from which
radio emission from the non-relativistic, spherical supernova ejecta
is still detected at late-time (\citealt{sks+05,bb05}, but see
\citealt{gr04}), none of the SNe Ibc show radio emission above our
average detection limit of $\sim 0.15$ mJy ($3\sigma$). In comparison
with SN\,1998bw, only SN\,2001em shows a comparable radio luminosity
on this timescale.  These results are consistent with the earlier
report by \citet{svw+03}.

\begin{figure}
\plotone{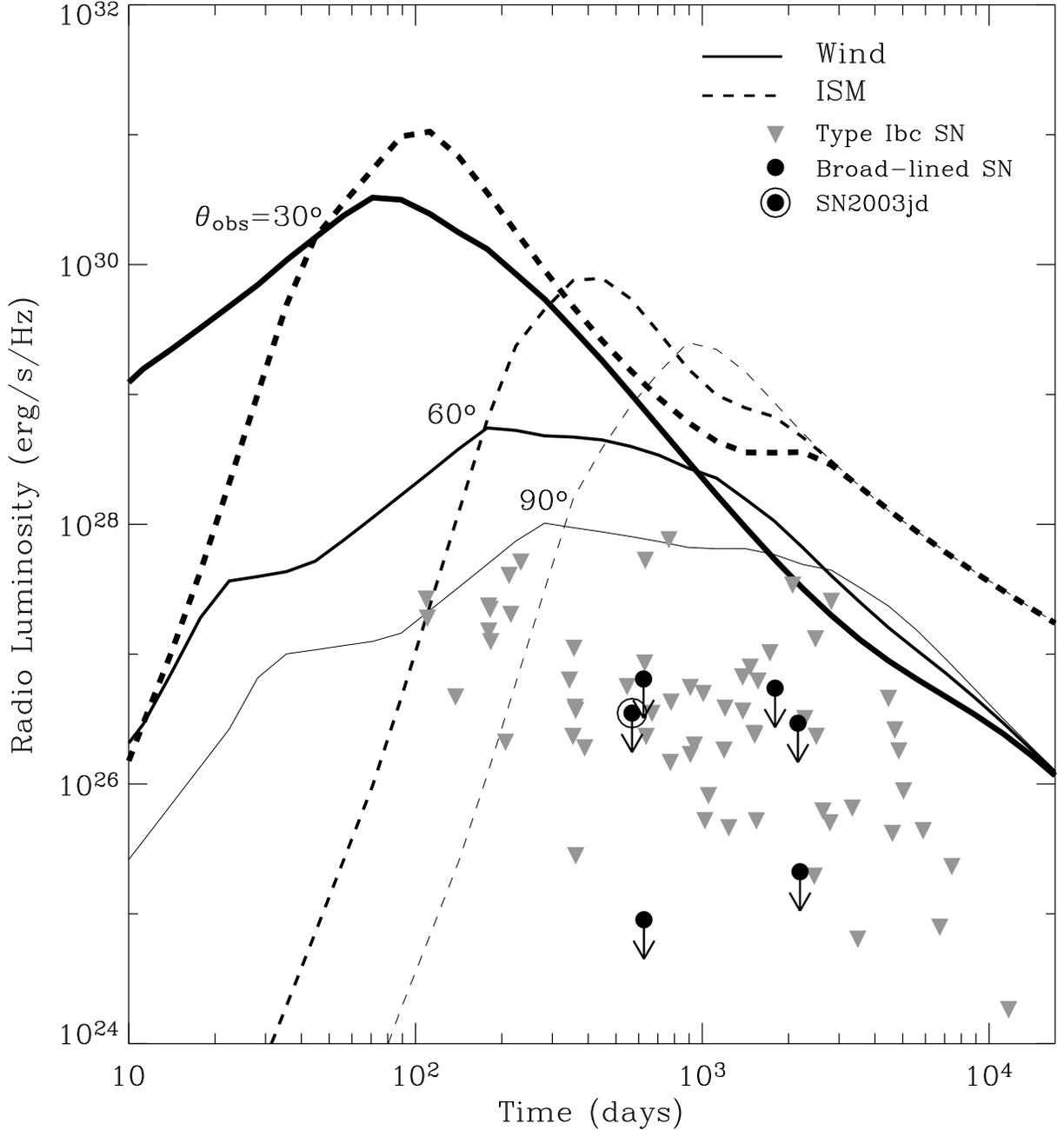}
\vspace{-0.8cm}
\caption{Late-time radio limits ($3\sigma$) are shown for 62 local
  Type Ibc supernovae (grey triangles) and six broad-lined SNe
  (circles/arrows) including SN\,2003jd (encircled dot).  For
  SN\,2001em, we adopt the spherical SN emission as an upper limit on
  the emission from an off-axis GRB jet.  Afterglow models for a
  typical GRB ($E_{51}=A_*=n_0=1$, $\epsilon_e=\epsilon_B=0.1$,
  $\theta_j=5^{\rm o}$, $p=2.2$) are shown for both wind-stratified
  (solid lines) and homogeneous (dashed lines) media at viewing angles
  of $\theta_{\rm obs}=30^{\rm o}$ (thickest), $60^{\rm o}$ (thick)
  and $90^{\rm o}$ (thin) away from the initial collimation axis of
  the jet.  As discussed in \S\ref{sec:our_model}, we make the
  conservative assumption that lateral jet spreading does not begin until
  after the non-relativistic transition, at which point the jets
  spread sideways at the sound speed.  All of the broad-lined SNe,
  including SN\,2003jd, are significantly fainter than the model
  light-curves, regardless of the CSM density profile or viewing
  angle.}
\label{fig:lum_limits_oa}
\end{figure}

In Figure~\ref{fig:lum_limits_oa} we plot the radio observations
for this sample of SNe Ibc, in addition to late-time radio data for
SN1954A \citep{ecb02} and SN\,1984L \citep{sfw04}.

\section{Off-Axis Models for Gamma-ray Bursts}
\label{sec:model}

\subsection{An Analytic Approach}
\label{sec:waxman}
\citet{wax04} present an analytic model for the late-time radio
emission from a typical GRB viewed significantly away from the
collimation axis.  In this model, the GRB jet is initially
characterized by a narrow opening angle, $\theta_j\sim$ few degrees,
while the viewing angle is assumed to be large, $\theta_{\rm
obs}\gtrsim 1$ radian.  As the jet sweeps up circumstellar material
(CSM) and decelerates, it eventually undergoes a dynamical transition
to sub-relativistic expansion \citep{fwk00}.  The timescale for this
non-relativistic transition is estimated at $t_{NR}\approx 0.2
(E_{51}/n_0)^{1/3}$ yr ($\approx 0.3 E_{51}/A_*$ yr) in the case of a
homogeneous (wind-stratified) medium \citep{wax04}.  Here, $E_{51}$ is
the beaming-corrected ejecta energy normalized to $10^{51}$ erg and
$n_0$ is the circumstellar density of the homogeneous medium
(interstellar medium; ISM) normalized to 1 particle cm$^{-3}$. For a
wind-stratified medium, $A_*$ defines the circumstellar density in
terms of the progenitor mass loss rate, $\dot{M}$, and wind velocity,
$v_w$, such that $\dot{M}/4\pi v_w=5\times 10^{11} A_*~\rm g~cm^{-1}$,
and thus $A_*=1$ for $\dot{M}=10^{-5}~M_{\odot}~\rm yr^{-1}$ and
$v_w=10^{3}~\rm km~s^{-1}$ \citep{lc99}.

Once sub-relativistic, the jets spread sideways, rapidly intersecting
our line-of-sight as the ejecta approach spherical symmetry.  At this
point the afterglow emission is effectively isotropic and
appears similar to both on-axis and off-axis observers.  The broadband
emission observed from the sub-relativistic ejecta is described by a
standard synchrotron spectrum, characterized by three break
frequencies: the synchrotron self-absorption frequency, $\nu_a$, the
characteristic synchrotron frequency, $\nu_m$, and the synchrotron
cooling frequency, $\nu_c$.  On timescales comparable to the
non-relativistic transition, $\nu_a$ and $\nu_m$ are typically below
the radio band while $\nu_c$ is generally near the optical
\citep{fwk00,bkf04,fsk+05}.  Making the usual assumption that the kinetic
energy is partitioned between relativistic electrons and magnetic
fields ($\epsilon_e$ and $\epsilon_B$, respectively), and that these
fractions are constant throughout the evolution of the jet,
\citet{wax04} estimate the radio luminosity of the sub-relativistic,
isotropic emission to be

\begin{eqnarray}
L_{\nu} & \approx & 8.0\times 10^{29} \left(\frac{\epsilon_e}{0.1}\right) \left(\frac{\epsilon_B}{0.1}\right)^{3/4} n_0^{3/4} E_{51} \\
& & \times \left(\frac{\nu}{\rm 10 GHz}\right)^{-1/2} \left(\frac{t}{t_{\rm NR}}\right)^{-9/10}~\rm erg~s^{-1}~Hz^{-1} \nonumber
\end{eqnarray}

\noindent
for the ISM case, while for a wind-stratified medium
 
\begin{eqnarray}
L_{\nu} & \approx & 2.1\times 10^{29} \left(\frac{\epsilon_e}{0.1}\right) \left(\frac{\epsilon_B}{0.1}\right)^{3/4} A_*^{9/4} E_{51}^{-1/2} \\
& & \times \left(\frac{\nu}{\rm 10 GHz}\right)^{-(p-1)/2} \left(\frac{t}{t_{\rm NR}}\right)^{-3/2}~\rm erg~s^{-1}~Hz^{-1}. \nonumber
\end{eqnarray}

\noindent
Here it is assumed that the electrons are accelerated
into a power-law distribution, $N(\gamma) \propto \gamma^{-p}$ with
$p=2.0$.  These equations reveal that the strength of the
non-relativistic emission is strongly dependent on the density of the
circumstellar medium (especially in the case of a wind) and is best
probed at low frequencies.

While this analytic model provides robust predictions for the
afterglow emission at $t>t_{\rm NR}$, it does not describe the early
evolution or the transition from relativistic to sub-relativistic
expansion.  At early time, the observed emission from an off-axis GRB
is strongly dependent on the viewing angle and dynamics of the jet.
To investigate this early afterglow evolution and the transition to
sub-relativistic expansion, we developed a detailed semi-analytic
model, described below.

\subsection{A Semi-analytic Model}
\label{sec:our_model}
In modeling the afterglow emission from an off-axis GRB jet, we adopt
the standard framework for a adiabatic blastwave expanding into either
a uniform or wind stratified medium \citep{sar97,gs02}.  We assume a
uniform, sharp-edged jet such that Lorentz factor and energy are
constant over the jet surface.  The hydrodynamic evolution of the jet
is fully described in \citet{onp04}. As the bulk Lorentz factor of the
ejecta approaches $\Gamma\sim 1$, the jets begin to spread laterally
at the sound speed \footnotemark\footnotetext{Since the spreading
  behavior of relativistic GRB jets is poorly constrained by
  observations, we assume negligible spreading during this phase.  We
  adopt this conservative assumption since it produces the faintest
  off-axis light-curves.}.  Our off-axis light-curves are obtained by
integrating the afterglow emission over equal arrival time surface.
We note that these resulting light-curves are in broad agreement with
Model 2 of \citet{gpk+02} and are consistent with Waxman's analytic
model (\S\ref{sec:waxman}) on timescales, $t\gtrsim t_{\rm NR}$.
 
Over-plotted in Figure~\ref{fig:lum_limits_oa} are our off-axis models
calculated for both wind-stratified and homogeneous media at an
observing frequency of $\nu_{\rm obs}=8.46$ GHz.  We assume standard
GRB parameters of $E_{51}=A_*=n=1$, $\epsilon_B=\epsilon_e=0.1$,
$p=2.2$ and $\theta_j=5^{\rm o}$, consistent with the typical values
inferred from broadband modeling of GRBs \citep{pk02,yhs+03,clf04}. We
compute model light-curves for off-axis viewing angles between 30 and
90 degrees.  As clearly shown in the figure, the majority of our
late-time SNe Ibc limits are significantly fainter than {\it all} of
the model light-curves, constraining even the extreme case where
$\theta_{\rm obs}=90^{\rm o}$.

\section{SN\,2003jd: Constraints on the off-axis jet}
\label{sec:SN2003jd}
Based on the double-peaked profiles observed for the nebular lines of
neutral oxygen and magnesium, \citet{mkm+05} argue that SN\,2003jd was
an aspherical, axisymmetric explosion viewed near the equatorial
plane. They suggest that this asymmetry may be explained if the SN
explosion was accompanied by a tightly collimated and relativistic GRB
jet, initially directed $\sim 70$ degrees from our line-of-sight.
This hypothesis is consistent with the observed lack of prompt
gamma-ray emission \citep{hcm+03} as well as the absence of strong
radio and X-ray emission at early time \citep{skf03,wpr+03}.

Our radio observation of SN\,2003jd at $t\sim 1.6$ years imposes
strong constraints on the putative off-axis GRB jet.  While the early
data constrain only mildly off-axis jets ($\theta_{\rm obs}\lesssim
30^{\rm o}$), our late-time epoch constrains even those jets initially
directed perpendicular to our line-of-sight.  As shown in
Figure~\ref{fig:lum_limits_oa}, our radio limit is a factor of
$\gtrsim 200$ ($\gtrsim 20$) fainter than that predicted for a typical
GRB expanding into a homogeneous (wind-stratified) medium, even in the
extreme case where $\theta_{\rm obs}\sim 90^{\rm o}$.  Given the
assumption of typical GRB parameters, we conclude that our late-time
radio limit is inconsistent with the presence of an off-axis GRB jet.
We note that the model assumptions and physical parameters of our
off-axis afterglow light-curves are identical to those adopted by
\citet{mkm+05}.

We next explore the range of parameters ruled out by our deep radio
limits.  As shown in Equations~1 and 2, the luminosity of the
late-time emission is a function of the ejecta energy, the density of
the circumstellar medium and the equipartition fractions.  To
investigate the effect of energy and density on the late-time radio
luminosity, we fix the equipartition fractions to
$\epsilon_e=\epsilon_B=0.1$, chosen to be consistent with the values
typically inferred from afterglow modeling of cosmological GRBs
\citep{pk02,yhs+03}.

\begin{figure}
\plotone{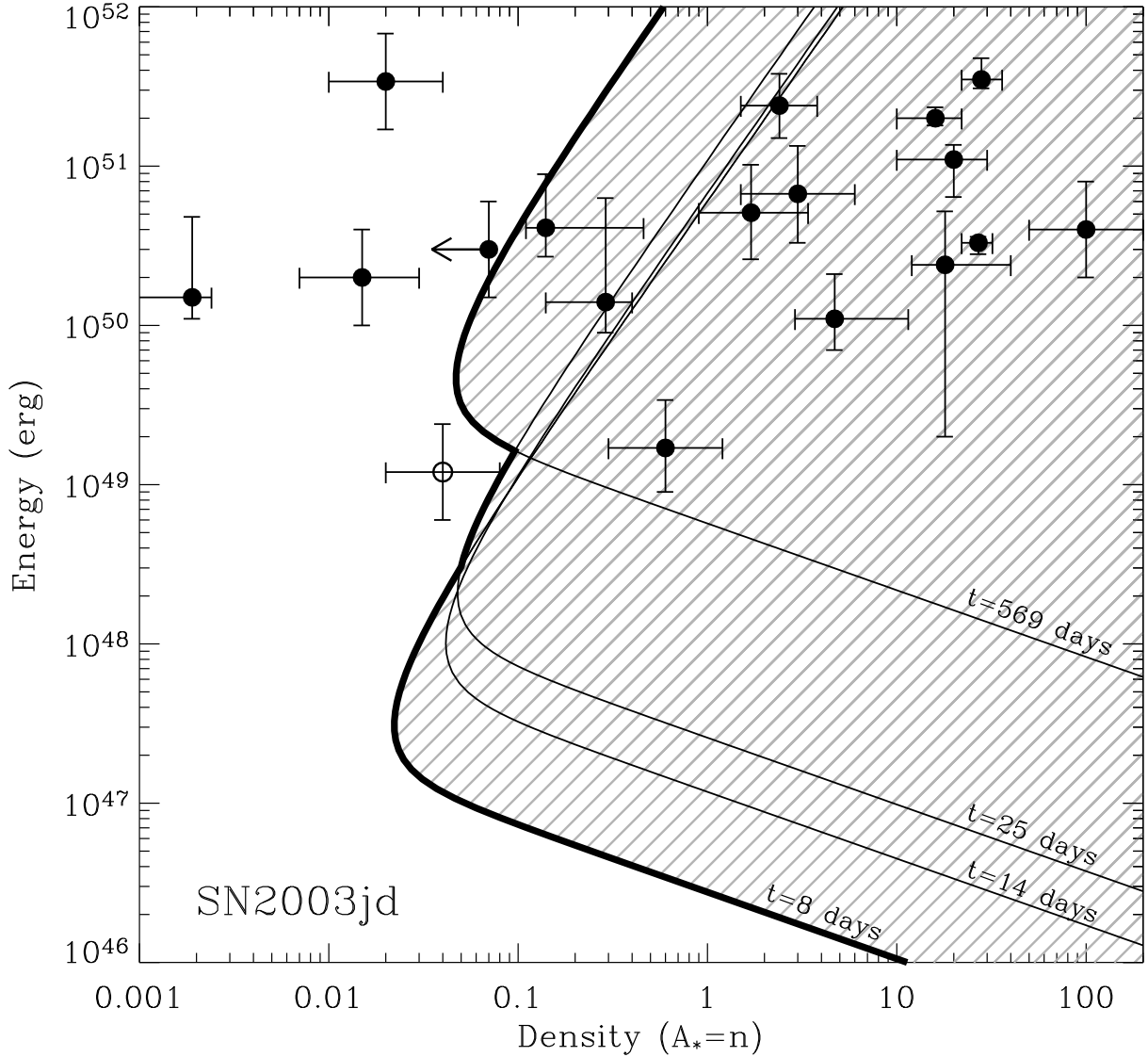}
\vspace{-0.5cm}
\caption{Constraints on an off-axis GRB associated with SN\,2003jd.
  Each of the four 8.46 GHz observations is mapped to a contour (thin
  black lines) in the two-dimensional parameter space of kinetic
  energy and CSM density for an off-axis GRB.  As discussed in
  \S\ref{sec:SN2003jd}, we adopt a conservative off-axis model
  (wind-stratified CSM, $\epsilon_e=\epsilon_B=0.1$, $p=2.2$,
  $\theta_{\rm obs}=90^{\rm o}$) which produces the faintest
  light-curves in Figure~1. We exclude the grey hatched region of
  parameter space {\it rightward} of each contour since this region
  produces a jet which is {\it brighter} than the observed $3\sigma$
  limit.  The union of these regions (thick black line) represents the
  total parameter space ruled out for an associated GRB.  For
  comparison, we also show are the energy and density values for 17
  GRBs (filled circles) from Table~\ref{tab:grb}, inferred from
  broadband modeling of the afterglow emission.  For the GRBs without
  reported error estimates, we adopt a factor of two for the
  uncertainty of both parameters.  We also include the parameters for
  GRB\,980425 (unfilled circle) derived from the early radio emission
  emitted from the quasi-spherical SN component, but note that the
  lack of strong radio emission from SN\,2003jd at early time
  indicates that it is not similar to GRB\,980425.}
\label{fig:SN2003jd_ed}
\end{figure}

In Figure~\ref{fig:SN2003jd_ed}, we illustrate how each radio epoch
for SN\,2003jd maps to a curve within the two-dimensional parameter
space of kinetic energy and circumstellar density for an off-axis GRB.
Here we adopt our semi-analytic model (\S\ref{sec:our_model})
for a wind-stratified medium, along with a typical electron index of
$p=2.2$ and a viewing angle of $\theta_{\rm obs}=90^{\rm o}$; the
faintest model for a given set of
equipartition fractions.  By comparing the luminosity limit for
SN\,2003jd at a particular epoch with the off-axis model prediction
for that time, we exclude the region of parameter space {\it
  rightward} of the curve since this region produces a jet which is
{\it brighter} than the observed limit.  The union of these regions
represents the total parameter space ruled out for an associated GRB.
As shown in this figure, the total excluded parameter space
extends from $A_*\gtrsim 0.03$ and $E \sim 10^{47}$ to $10^{52}$.

We compare these constraints with the beaming-corrected kinetic
energies and CSM densities for 18 cosmological GRBs
(Table~\ref{tab:grb}).  Here we make the rough approximation that
$A_*\approx n_0$; a reasonable assumption for circumstellar radii near
$\sim 10^{18}$ cm. As shown in Figure~\ref{fig:SN2003jd_ed}, these
GRBs span the region of parameter space roughly bracketed by $A_*\sim
0.002$ to 100 and $E\sim 2\times 10^{49}$ to $4\times 10^{51}$. The majority of the bursts
(13 out of 18) fall within the excluded region of parameter space for
SN\,2003jd.  We conclude that SN\,2003jd was not likely associated
with a typical GRB at a confidence level of $\sim 72\%$.

\section{Local Type Ibc Supernovae: Further Constraints}
\label{sec:SNe}

\begin{figure}
\plotone{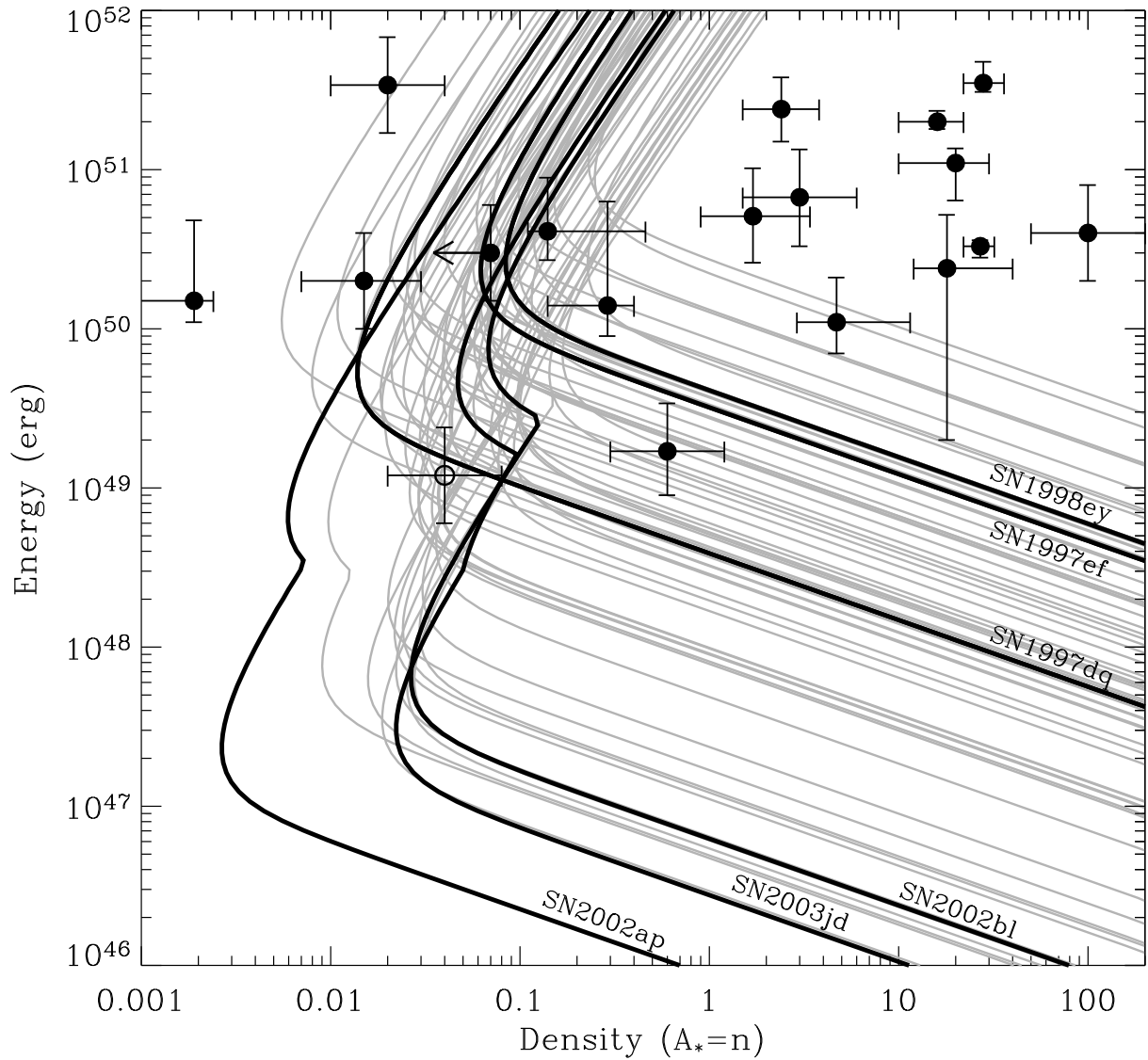}
\caption{Late-time radio limits for 62 SNe Ibc (grey contours) and six
  broad-lined SNe (black contours, labels) are used to constrain the
  parameters ($E_{51}$ and $A_*$) of putative off-axis GRBs as
  discussed in \S\ref{sec:SNe}.  For SN\,2001em, we adopt
  the spherical SN emission as an upper limit on the emission from an
  off-axis GRB jet.  As discussed in \S\ref{sec:SN2003jd}, we adopt a
  conservative off-axis model (wind-stratified CSM,
  $\epsilon_e=\epsilon_B=0.1$, $p=2.2$, $\theta_{\rm obs}=90^{\rm o}$)
  which produces the faintest light-curves in
  Figure~1.  The region rightward of each SN contour is ruled out
  since it implies an off-axis GRB which is brighter than the observed
  $3\sigma$ limit. For the 21 SNe with additional early-time radio
  observations, we determine the union of parameter space which is
  excluded (see Figure~\ref{fig:SN2003jd_ed}). Energy and density
  values for 17 cosmological GRBs (filled circles) and GRB\,980425
  (unfilled circle) are shown for comparison.  For the GRBs without
  reported error estimates, we adopt a factor of two for the
  uncertainty of both parameters.  It is clear that most of the GRBs
  lie in the excluded regions for each SN, implying that if these
  events are associated with off-axis GRBs, their densities are
  unusually low.}
\label{fig:all_ed}
\end{figure}

While physical parameters atypical of the cosmological GRB population
can be invoked to hide an off-axis GRB for SN\,2003jd, it is
exceedingly unlikely for atypical parameters to dominate a large
statistical sample of SNe Ibc.  Motivated thus, we searched for
off-axis GRBs in the 67 local Type Ibc SNe for which we have compiled
late-time ($t\sim 0.5-30$ yr) radio observations.  Applying the method
described in \S\ref{sec:SN2003jd} we produce exclusion regions in the
$E_{51}-A_*$ parameter space for each SN.  Figure~\ref{fig:all_ed}
shows the resulting contours for all 68 SNe, including SN\,2003jd and
five broad-lined events.  For the twenty SNe with early radio limits
\citep{bkf+03,bkc02} we combine late- and early-time data to provide
further constraints.

\begin{figure}
\plotone{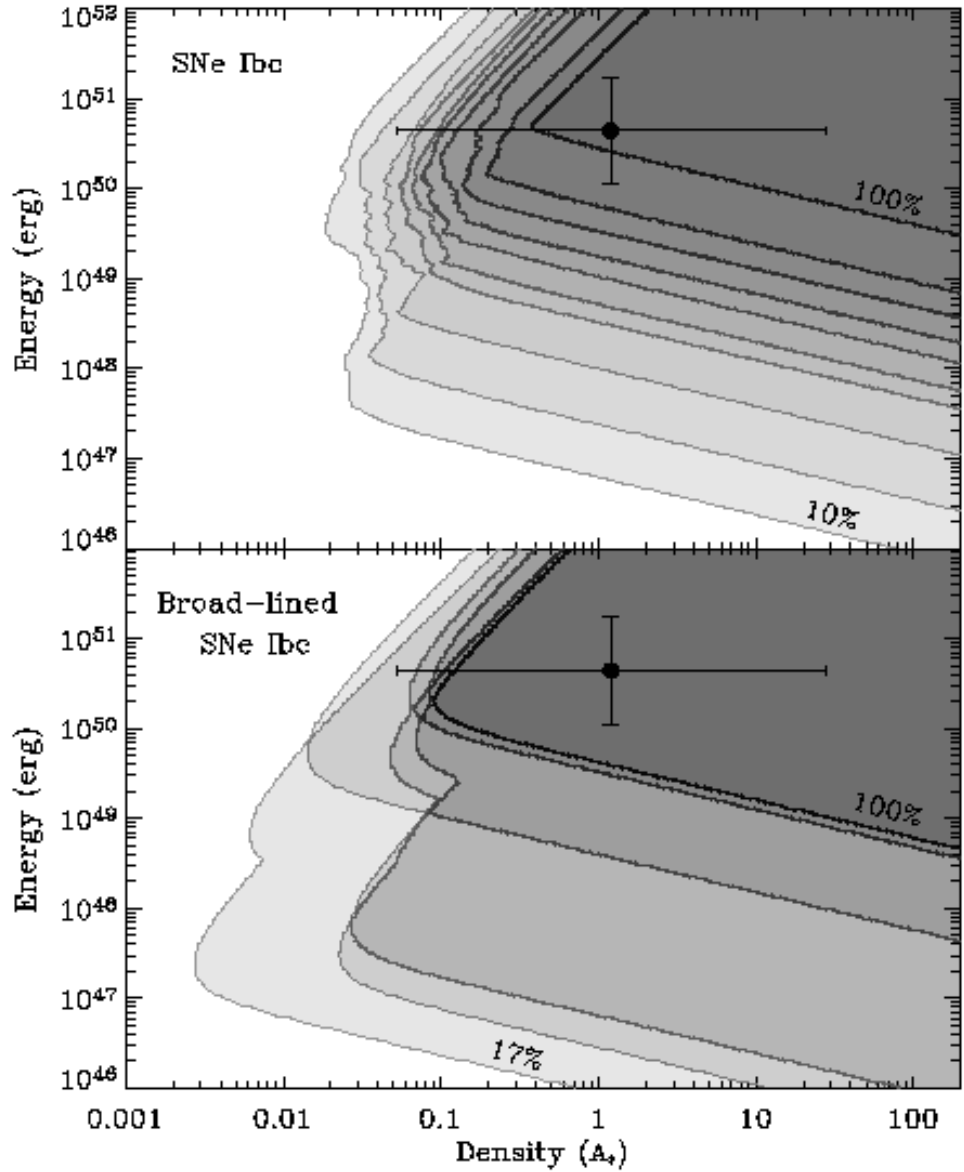}
\vspace{-2.5cm}
\caption{Top panel: Regions of $E_{51}-A_*$ parameter space for
  off-axis GRBs ruled out by the SNe Ibc sample in
  Figure~\ref{fig:all_ed}. Contours/shading depict the regions ruled
  out by fractions of the sample, labeled in increments of 10\%.  The
  darkest shading corresponds to the region excluded by 100\% of the
  sample. The statistical mean $E_{51}$ and $A_*$ values for
  cosmological GRBs is shown as a barred point. Bottom panel: Same as
  above, except only including the broad-lined SNe.}
\label{fig:confidence_ed}
\end{figure}

In Figure~\ref{fig:confidence_ed} we compile all 68 exclusion regions
to quantify the $E_{51}-A_*$ parameter space constrained by this
statistical sample.  Contours map the regions excluded by incremental
fractions of our sample.  As in the case of SN\,2003jd, all curves
rule out bursts with $A_*\gtrsim 1$ and $E\gtrsim 10^{50}$ erg.
Moreover, 50\% exclude $A_* \gtrsim 0.1$ and $E \gtrsim 10^{49}$ erg.
For comparison, the mean ejecta energy and CSM density values for cosmological
GRBs are $E \approx 4.4\times 10^{50}$ erg and $A_*=n_0\approx 1.2$.

Focusing on the subsample of broad-lined SNe, we emphasize that our
deep limits rule out both putative GRB jets directed along our
line-of-sight (e.g. SN\,1997ef) as well as those which are initially
beamed off-axis (e.g. SN\,2002ap and SN\,2003jd).  In particular, the
large exclusion region for SN\,2002ap (see Figure~\ref{fig:all_ed})
implies that an extremely low CSM density, less than $A_*\sim 3\times
10^{-3}$, is needed to suppress the emission from an associated GRB.
This is a factor of $\sim 10$ below the density inferred from modeling
of the early radio emission \citep{bkc02} and we therefore conclude
that an off-axis GRB model is inconsistent with our late-time
observations of SN\,2002ap.  In Figure~\ref{fig:confidence_ed} we show
that this entire sample of six broad-lined SNe rule out bursts with
energies $E\gtrsim 10^{49}$ erg, and 50\% even rule out $E\sim 10^{47}$ erg
(all assuming a typical $A_*=1$).

\begin{figure}
\plotone{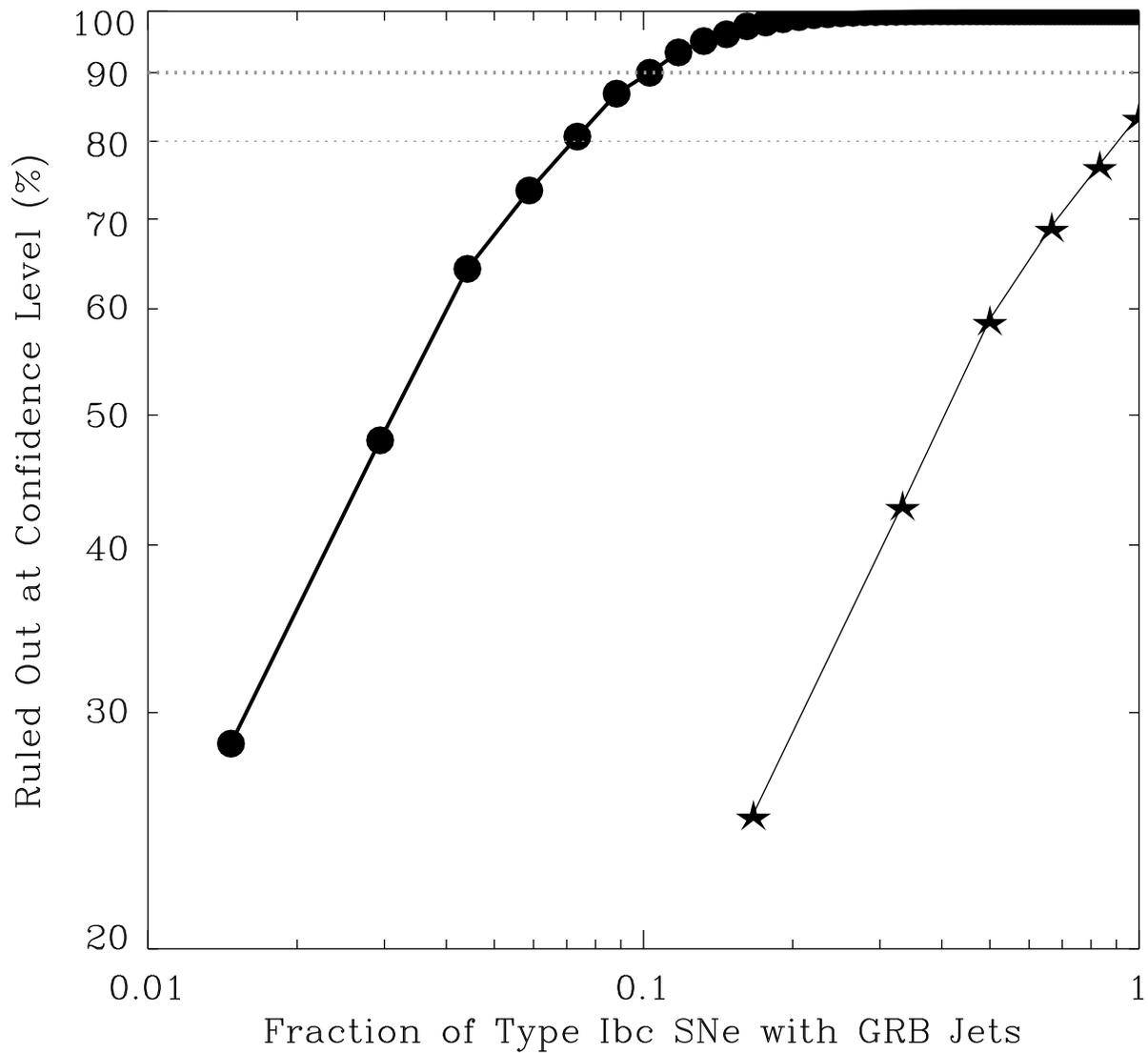}
\caption{Using late-time radio data for 68 local SNe Ibc
  (Table~\ref{tab:vla}) and the physical parameters inferred for
  cosmological GRBs (Table~\ref{tab:grb}) we constrain the fraction of
  SNe harboring GRBs.  We find that we can rule out the the scenario
  in which every supernova (circles) has an associated GRB
  (GRBs/SNe=1) at a confidence level of $\sim 100\%$ and can rule out
  a fraction of $\gtrsim 0.1$ with a confidence level of $\sim 90\%$.
  For the broad-lined supernovae (stars) we rule out the scenario in
  which every event harbors a GRB with a confidence level of $\sim
  84\%$.}
\label{fig:binomial_prob}
\end{figure}

We next address the limits on an association with GRBs defined by the
cosmological sample (Table~\ref{tab:grb}).  For each SN in our sample
we calculate the fraction of observed GRBs that lie in its exclusion
region. We then determine the probability of finding null-detections
for our entire sample by calculating the product of the individual
probabilities.  We find that the probability that {\it every} Type Ibc
supernova has an associated GRB is $1.1\times 10^{-10}$.  We further
rule out a scenario in which one in ten SNe Ibc is associated with a
GRB at a confidence level of $\sim 90\%$.  For the broad-lined events
alone we rule out the scenario that every event is associated with a
GRB at a confidence level of $\sim 84\%$.  Confidence levels are shown
as a function of GRB/SN fraction in Figure~\ref{fig:binomial_prob}.

\section{Discussion and Conclusions}
\label{sec:disc}

We present late-time radio observations for 68 local Type Ibc supernovae,
including six broad-lined SNe (``hypernovae''), making this the most
comprehensive study of late-time radio emission from SNe Ibc. None of
these objects show evidence for bright, late-time radio emission that
could be attributed to off-axis jets coming into our line-of-sight.
Comparison with our most conservative off-axis GRB afterglow models
reveals the following conclusions:

(1) Less than $\sim 10\%$ of Type Ibc supernovae are associated with
GRBs.  These data impose an empirical constraint on the GRB beaming
factor, $\langle f_b^{-1}\rangle$, where $f_b=(1-{\rm cos}~\theta_j)$.
Assuming a local GRB rate of $\sim 0.5~\rm Gpc^{-3}~yr^{-1}$
\citep{sch01,psf03,gpw05} and an observed SNe Ibc rate of $\sim
4.8\times 10^{4}~\rm Gpc^{-3}~yr^{-1}$ \citep{mdp+98,cet99,frp+99}, we
constrain the GRB beaming factor to be $\langle
f_b^{-1}\rangle\lesssim \times 10^{4}$.  Adopting a lower limit of
$\langle f_b^{-1}\rangle > 13$ \citep{low+02}, the beaming factor is
now observationally bound by $\langle f_b^{-1}\rangle\approx
[13-10^4]$, consistent with the observed distribution of jet
opening angles \citep{fks+01,gpw05}.

(2) Despite predictions that most or all broad-lined SNe Ibc harbor
GRB jets \citep{pmn+04}, our radio observations for six broad-lined
events (SNe 1997dq, 1997ef, 1998ey, 2002ap, 2002bl and 2003jd) reveal
no evidence for association with typical (or even sub-energetic) GRBs.
While unusual physical parameters can suppress the radio emission from
off-axis jets in any one SN, it is unlikely that all six broad-lined
events host atypical GRBs.  We observationally rule out the scenario
in which every broad-lined SN harbors GRB jets with a confidence level
of $\sim 84\%$.

(3) While low CSM densities (e.g. $A_* \lesssim 0.1$) can suppress the
emission from off-axis GRB jets, such values are inconsistent
with the mass loss rates measured from local Wolf-Rayet stars
($0.6-9.5\times 10^{-5}~\rm M_{\odot}~yr^{-1}$; \citealt{cgv04}),
thought to be the progenitors of long-duration gamma-ray bursts.

(4) While we have so far considered only the signature from a highly
collimated GRB jet, these late-time radio data also impose constraints
on the presence of broader jets and/or jet cocoons.  As demonstrated
by GRBs 980425 and 030329, the fraction of energy coupled to mildly
relativistic and mildly collimated ejecta can dominate the total
relativistic energy budget \citep{kfw+98,bkp+03}.  Less sensitive to
to the effects of beaming and viewing geometry, broad jets are more
easily probed at early time ($t\sim 100$ days) when the emission is
brightest.  Still, we note that the majority of our late-time radio
limits are significantly fainter than GRBs 980425 and 030329 on a
comparable timescale, thus constraining even mildly relativistic
ejecta.

These conclusions, taken together with the broad spectral features
observed for GRB-associated SNe 1998bw, 2003dh and 2003lw, motivate
the question: what is the connection between GRBs and local Type Ibc
supernovae?  While current optical data suggest that all GRB-SNe are
broad-lined, our late-time radio observations clearly show that the
inverse is {\em not} true: broad optical absorption lines do not
serve as a reliable proxy for relativistic ejecta. This suggests that
their observed large photospheric velocities and asymmetric
ejecta are often merely characteristics of the non-relativistic SN explosion
and thus manifestations of the diversity within SNe Ibc.

The authors thank Doug Leonard, Paolo Mazzali, Dale Frail, Brian
Schmidt and Avishay Gal-Yam for helpful discussions.  As always, the
authors thank Jochen Greiner for maintaining his GRB page.  A.M.S. is
supported by the NASA Graduate Student Research Program.  E.B. is
supported by NASA through Hubble Fellowship grant HST-HF-01171.01
awarded by the STScI, which is operated by the Association of
Universities for Research in Astronomy, Inc., for NASA, under contract
NAS 5-26555.

\bibliographystyle{apj1b}

\clearpage

\begin{deluxetable}{lrrrr}
\tablecaption{Radio Observations of SN\,2003jd}
\tablewidth{0pt} \tablehead{
\colhead{Date Obs} & \colhead{$\Delta t$\tablenotemark{a}} & \colhead{$F_{\nu,4.96~\rm GHz}$\tablenotemark{b}} & \colhead{$F_{\nu,8.46~\rm GHz}$} & \colhead{$F_{\nu,22.5~\rm GHz}$} \\
\colhead{(UT)} & \colhead{(days)} & \colhead{($\mu$Jy)} & \colhead{($\mu$Jy)} & \colhead{($\mu$Jy)}
}
\startdata
2003 Oct 29 & 8 & $\pm 52$ & $\pm 34$ & $\pm 58$ \\
2003 Nov 4 & 14 & \nodata & $\pm 77$ & \nodata \\
2003 Nov 15 & 25 & \nodata & $\pm 74$ & \nodata \\
2005 May 12 & 569 & \nodata & $\pm 15$ & \nodata \\
\enddata
\tablenotetext{a}{Assuming an explosion date of 2003 October 21 UT, based on pre-explosion images \citep{bsl+03}.}  
\tablenotetext{b}{All flux densities are given as $1\sigma$ (rms).}  
\label{tab:SN2003jd}
\end{deluxetable}

\clearpage

\begin{deluxetable}{llrcclrr}
\tablecaption{Late-time Radio Observations of Type Ibc Supernovae}
\tablewidth{0pt} \tablehead{
\colhead{SN name} & \colhead{Host Galaxy} & \colhead{Distance\tablenotemark{a}} & \colhead{Explosion Date\tablenotemark{b}} & \colhead{IAUC} & \colhead{Date Obs} & \colhead{$\Delta t$\tablenotemark{c}} & \colhead{Flux density\tablenotemark{d}} \\
\colhead{} & \colhead{}  & \colhead{(Mpc)} & \colhead{(UT)} & \colhead{(\#)} & \colhead{(UT)} & \colhead{(days)} & \colhead{($\mu$Jy)}
}
\startdata  
1983N  & NGC 5236         & 7.2       & 1983 Jun 26                & 3835 & 2003 Oct 17           & 7416             & $\pm 124$ \\       
1985F  & NGC 4618         & 7.7       & 1985 May 14               & 4042 & 2003 Oct 17           & 6730             & $\pm 37$ \\     
1987M  & NGC 2715         & 18.9      & 1987 Aug 31               & 4451 & 2003 Oct 17           & 5891             & $\pm 34$\\
1990B  & NGC 4568         & 32.0      & 1989 Dec 23-1990 Jan 20   & 4949 & 2003 Oct 17           & 5032             & $\pm 24$ \\
1990U  & NGC 7479         & 33.7      & 1990 Jun 28-Jul 27        & 5063 & 2003 Oct 17           & 4845             & $\pm 44$ \\       
1991A  & IC 2973          & 45.5      & 1990 Dec 6-13             & 5178 & 2003 Oct 17           & 4695             & $\pm 35$ \\        
1991D  & Anon.            & 173       & 1991 Jan 16               & 5153 & 1992 Oct 11           & 633             & $\pm 49$ \\
1991N  & NGC 3310         & 14.0      & 1991 Feb 20-Mar 29        & 5227 & 2003 Oct 17           & 4604             & $\pm 59$ \\        
1991ar & IC 49            & 65.0      & 1991 Aug 3                & 5334 & 2003 Oct 17           & 4458             & $\pm 30$ \\
1994I  & NGC 5194         & 6.5       & 1994 Apr 1-2              & 5961 & 2003 Oct 17           & 3486             & $\pm 42$ \\
1994ai & NGC 908          & 21.3      & 1994 Dec 8-16             & 6120 & 2003 Oct 17           & 3336             & $\pm 40$ \\
1996D  & NGC 1614         & 68.1      & 1996 Jan 28               & 6317 & 2003 Oct 15           & 2818             & $\pm 153$ \\
1996N  & NGC 1398         & 19.7      & 1996 Feb 16-Mar 13        & 6351 & 2003 Oct 15           & 2787             & $\pm 36$ \\
1996aq & NGC 5584         & 23.2      & 1996 Jul 30               & 6465 & 2003 Oct 17           & 2635             & $\pm 32$ \\
1997B  & IC 438           & 44.3      & 1996 Dec 14               & 6535 & 2003 Oct 15           & 2498             & $\pm 33$ \\
1997C  & NGC 3160         & 99.2      & 1996 Dec 18-1997 Jan 14   & 6536 & 2003 Oct 17           & 2481             & $\pm 37$ \\
1997X  & NGC 4691         & 15.7      & 1997 Jan 16-Feb 2         & 6552 & 2003 Oct 17           & 2457             & $\pm 22$ \\
1997dc & NGC 7678         & 49.6      & 1997 Jul 22               & 6715 & 2003 Oct 17           & 2278             & $\pm 36$ \\
1997dq\tablenotemark{\star} & NGC 3810         & 14.0      & 1997 Oct 16  & 6770              & 2003 Oct 17           & 2192             & $\pm 30$ \\
1997ef\tablenotemark{\star} & UGC 4107         & 49.8      & 1997 Nov 16-26    & 6778        & 2003 Oct 17           & 2156             & $\pm 33$ \\
1998T  & NGC 3690         & 44.3      & 1998 Feb 8-Mar 3          & 6830 & 2003 Oct 17         & 2066             & $\pm 482$ \\
1998ey\tablenotemark{\star} & NGC 7080         & 69.0      & 1998 Nov 1-Dec 5  & 6830          & 2003 Oct 17           & 1794             & $\pm 32$ \\
1999bc & UGC 4433         & 90.2      & 1999 Jan 30               & 7133 & 2003 Oct 15           & 1721             & $\pm 25$ \\
1999di & NGC 776          & 70.2      & 1999 Jul 6                & 7234 & 2003 Oct 17           & 1564             & $\pm 35$ \\
1999dn & NGC 7714         & 39.7      & 1999 Aug 10-20            & 7241 & 2003 Oct 17           & 1524             & $\pm 42$ \\
1999ec & NGC 2207         & 38.9      & 1999 Aug 24               & 7268 & 2003 Oct 15           & 1515             & $\pm 45$ \\
1999eh & NGC 2770         & 27.6      & 1999 Jul 26               & 7282 & 2003 Oct 17           & 1544             & $\pm 19$ \\
2000C\tablenotemark{\dagger}  & NGC 2415         & 53.8      & 1999 Dec 30-2000 Jan 4 & 7348   & 2003 Oct 17           & 1385             & $\pm 35$ \\
2000F  & IC 302           & 84.5      & 1999 Dec 30-2000 Jan 10   & 7353 & 2003 Oct 15           & 1383             & $\pm 26$ \\
2000S  & MCG -01-27-2     & 85.8      & 1999 Oct 9                & 7384 & 2003 Oct 17           & 1469             & $\pm 30$ \\
2000de & NGC 4384         & 35.6      & 2000 Jul 13               & 7478 & 2003 Oct 17           & 1191             & $\pm 40$\tablenotemark{e} \\
2000ds & NGC 2768         & 19.4      & 2000 May 28               & 7507 & 2003 Oct 17           & 1237             & $\pm 34$ \\
2000dv & UGC 4671         & 57.7      & 2000 Jul 4                & 7510 & 2003 Oct 17           & 1200             & $\pm 32$ \\
2001B\tablenotemark{\dagger}  & IC 391           & 22.0      & 2000 Dec 25-2001 Jan 4  & 7555  & 2003 Oct 15           & 1021             & $\pm 30$ \\
2001M\tablenotemark{\dagger}  & NGC 3240         & 50.6      & 2001 Jan 3-21     & 7568         & 2003 Oct 17           & 1008             & $\pm 54$ \\
2001bb\tablenotemark{\dagger} & IC 4319          & 66.3      & 2001 Apr 15-29    & 7614         & 2003 Oct 17           & 908              & $\pm 35$ \\
2001ch & MCG -01-54-1     & 41.6      & 2001 Mar 23               & 7637       & 2003 Oct 17 & 938              & $\pm 32$ \\
2001ci\tablenotemark{\dagger} & NGC 3079         & 15.8      & 2001 Apr 17-25     & 7638       & 2003 Oct 17           & 909              & $\pm 188$ \\
2001ef\tablenotemark{\dagger} & IC 381           & 35.1      & 2001 Aug 29-Sep 9  & 7710       & 2003 Oct 17           & 774              & $\pm 33$ \\
2001ej\tablenotemark{\dagger} & UGC 3829         & 57.4      & 2001 Aug 30      & 7719         & 2003 Oct 15           & 778              & $\pm 36$ \\
2001em & UGC 11794        & 83.6      & 2001 Sep 10-15            & 7722 & 2003 Oct 17           & 765              & $907\pm 58$ \\
2001is\tablenotemark{\dagger} & NGC 1961         & 56.0      & 2001 Dec 14-23    & 7782         & 2003 Oct 15           & 668              & $\pm 31$ \\
2002J\tablenotemark{\dagger}  & NGC 3464         & 53.0      & 2002 Jan 15-21    & 7800         & 2003 Oct 17           & 637              & $\pm 23$ \\
2002ap\tablenotemark{\star\ddagger} & NGC 628          & 9.3       & 2002 Jan 28-29  & 7810          & 2003 Oct 17           & 626              & $\pm 29$ \\
2002bl\tablenotemark{\star\dagger} & UGC 5499         & 67.8      & 2002 Jan 31    & 7845           & 2003 Oct 17           & 624              & $\pm 39$ \\
2002bm\tablenotemark{\dagger} & MCG -01-32-1     & 78.0      & 2002 Jan 16-Mar 6   & 7845      & 2003 Oct 17           & 630              & $\pm 39$ \\
2002cp\tablenotemark{\dagger} & NGC 3074         & 73.4      & 2002 Apr 11-28    & 7887        & 2003 Oct 17           & 547              & $\pm 29$ \\
2002hf\tablenotemark{\dagger} & MCG -05-3-20     & 80.2      & 2002 Oct 22-29   & 8004         & 2003 Oct 17           & 357              & $\pm 48$ \\
2002ho\tablenotemark{\dagger} & NGC 4210         & 38.8      & 2002 Sep 24      & 8011         & 2003 Oct 17           & 389              & $\pm 35$ \\
2002hy\tablenotemark{\dagger} & NGC 3464         & 53.0      & 2002 Oct 13-Nov 12  & 8016      & 2003 Oct 17           & 354              & $\pm 23$ \\
\tablebreak
2002hz\tablenotemark{\dagger} & UGC 12044        & 77.8      & 2002 Nov 2-12      & 8017       & 2003 Oct 17           & 344              & $\pm 29$ \\
2002ji\tablenotemark{\dagger} & NGC 3655         & 20.8      & 2002 Oct 20      & 8025         & 2003 Oct 17           & 362              & $\pm 18$ \\
2002jj\tablenotemark{\dagger} & IC 340           & 60.1      & 2002 Oct 22      & 8026         & 2003 Oct 15           & 360              & $\pm 30$ \\ 
2002jp\tablenotemark{\dagger} & NGC 3313         & 52.7      & 2002 Oct 20     & 8031          & 2003 Oct 17           & 362              & $\pm 37$ \\
2003A  & UGC 5904         & 94.4      & 2002 Nov 23-2003 Jan 5    & 8041 & 2003 Jun 15           & 182              & $\pm 69$ \\
2003I  & IC 2481          & 76.0      & 2002 Dec 14               & 8046 & 2003 Jun 15           & 183              & $\pm 60$ \\
2003aa & NGC 3367         & 43.1      & 2003 Jan 24-31            & 8063 & 2003 Jun 15           & 138              & $\pm 70$ \\
2003bm & UGC 4226         & 113.7     & 2003 Feb 20-Mar 3         & 8086 & 2003 Jun 15           & 109              & $\pm 57$ \\
2003bu & NGC 5393         & 86.1      & 2003 Feb 10-Mar 11        & 8092 & 2003 Jun 15           & 110              & $\pm 71$ \\
2003cr & UGC 9639         & 155.2     & 2003 Feb 24-Mar 31        & 8103 & 2003 Nov 2            & 233              & $\pm 59$ \\
2003dg & UGC 6934         & 79.1      & 2003 Mar 24-Apr 8         & 8113 & 2003 Nov 2            & 215              & $\pm 90$ \\
2003dr & NGC 5714         & 31.7      & 2003 Apr 8-12             & 8117 & 2003 Nov 2            & 206              & $\pm 58$ \\
2003ds & NGC 3191         & 132.9     & 2003 Mar 25-Apr 14        & 8120 & 2003 Nov 2            & 212              & $\pm 67$ \\
2003el & NGC 5000         & 80.2      & 2003 Apr 19-May 22        & 8135 & 2003 Nov 2            & 180              & $\pm 65$ \\
2003ev & Anonymous        & 103.3     & 2003 Apr 9-Jun 1          & 8140 & 2003 Nov 2            & 180              & $\pm 61$ \\
2003jd\tablenotemark{\star\dagger} & MCG -01-59-21 & 80.8 & 2003 Oct 16-25 & 8232 & 2005 May 12       & 569              & $\pm 15$ \\
\enddata 
\tablenotetext{a}{Distances have been calculated from the
  host galaxy redshift.  We assume $H_0=72~\rm km~s^{-1}~Mpc^{-1}$,
  $\Omega_M=0.27$ and $\Omega_{\Lambda}=0.73$.}
\tablenotetext{b}{Explosion dates are given as a range constrained by
  the date of discovery and the most recent pre-discovery image in
  which the SN is not detected.  In the cases where pre-discovery
  images are not available (or not constraining), we adopt the
  spectroscopically derived age estimate with respect to maximum light
  and assume a typical light-curve rise time of $\sim 21$ days to
  provide a rough estimate of the explosion date.}
\tablenotetext{c}{Calculated using the average of the explosion date
  range in Column 4.}  
\tablenotetext{d}{All observations were conducted at 8.46 GHz (except for SN\,1991D at 4.86 GHz) and uncertainties represent $1\sigma$ rms noise.}  
\tablenotetext{e}{We detect a $\sim 0.7$ mJy source located $\sim 2$ arcsec from the optical position for SN\,2000de.  Due to the significant positional offset, we assume the radio source is not associated with the SN.}
\tablenotetext{\star}{Broad absorption lines observed spectroscopically.}
\tablenotetext{\dagger}{Early radio limits reported in \citet{bkf+03}}.
\tablenotetext{\ddagger}{Early radio data for SN\,2002ap was reported by \citet{bkc02}}.
\label{tab:vla}
\end{deluxetable}

\clearpage

\begin{deluxetable}{lrrrcc}
\tablecaption{Physical Parameters for GRBs}
\tablewidth{0pt} \tablehead{
\colhead{GRB} & \colhead{z} & \colhead{Ejecta Energy\tablenotemark{a}} & \colhead{Density\tablenotemark{b}} & \colhead{Density Profile} & \colhead{Reference\tablenotemark{c}} \\
\colhead{} & \colhead{} & \colhead{($\times 10^{51}$ erg)} & \colhead{($A_*=n$)} & \colhead{} & \colhead{}
}
\startdata
970508 & 0.835 & $2.4^{+1.4}_{-0.9}$ & $2.4^{+1.4}_{-0.9}$ & ISM & 1 \\
980329 & $\lesssim 3.9$ & $1.1^{+0.26}_{-0.46}$ & $20^{+10}_{-10}$ & ISM & 2 \\
980425 & 0.0085 & 0.012 & 0.04 & Wind & 3 \\
980519 & 1\tablenotemark{\dagger} & $0.41^{+0.48}_{-0.14}$ & $0.14^{+0.48}_{-0.14}$ & ISM & 4 \\
980703 & 0.966 & $3.5^{+1.26}_{-0.42}$ & $28^{+8}_{-6}$ & ISM & 2 \\
990123 & 1.60 & $0.15^{+0.33}_{-0.04}$ & $0.0019^{+0.0005}_{-0.0015}$ & ISM & 4 \\
990510 & 1.619 & $0.14^{0.49}_{-0.05}$ & $0.29^{+0.11}_{-0.15}$ & ISM & 4 \\
991208 & 0.706 & $0.24^{+0.28}_{-0.22}$ & $18^{+22}_{-6}$ & ISM & 4 \\
991216 & 1.02 & $0.11^{+0.1}_{-0.04}$ & $4.7^{+6.8}_{-1.8}$ & ISM & 4 \\
000301c & 2.03 & $0.33^{+0.03}_{-0.05}$ & $27^{+5}_{-5}$ & ISM & 4 \\
000418 & 1.118 & 3.4 & 0.02 & ISM & 5 \\
000926 & 2.066 & $2.0^{+0.34}_{-0.2}$ & $16^{+6}_{-6}$ & ISM & 2 \\
010222 & 1.477 & 0.51 & 1.7 & ISM & 4 \\
011121 & 0.36 & 0.2 & 0.015 & Wind & 6\\
020405 & 0.69 & 0.3 & $\lesssim 0.07$ & Wind & 6 \\
020903 & 0.251 & 0.4 & 100 & ISM & 7 \\
030329 & 0.168 & 0.67 & 3.0 & ISM & 8 \\
031203 & 0.105 & 0.017 & 0.6 & ISM & 9 \\
\enddata
\tablenotetext{a}{Corrected for beaming. Errors bracket the 90\% confidence interval.}
\tablenotetext{b}{Errors bracket the 90\% confidence interval.}
\tablenotetext{c}{References: 1. \citet{bkf04}; 2. \citet{yhs+03}; 3. \citet{lc99}; 4. \citet{pk02}; 5. \citet{bdf+01}; 6. \citet{clf04}; 7. \citet{skb+04a}; 8. \citet{bkp+03}; 9. \citet{skb+04b}}
\tablenotetext{\dagger}{Redshift Unknown; z=1 was assumed.}
\label{tab:grb}
\end{deluxetable}

\clearpage

\clearpage

\clearpage

\end{document}